\documentclass[12pt]{article}
\usepackage[pdftex]{graphicx}
\usepackage{wrapfig}
\usepackage{latexsym}
\usepackage[pdftex]{color}
\usepackage{pifont}
\usepackage{epstopdf}
\usepackage{tikz}
\usepackage{mathtools}
\usetikzlibrary{decorations.pathmorphing,patterns}

\newcommand{\bmp}[2][t]{\begin{minipage}[#1]{#2}}
\newcommand{\emp}{\end{minipage}}

\usepackage{amsmath}
\usepackage{amssymb}
\usepackage{textcomp}
\usepackage{tcolorbox}
\usepackage[title]{appendix}

\usepackage{sidecap,caption}

\usepackage[square,numbers,sort&compress]{natbib}

\newcommand\black{\color{black}}

\newcommand{\brev}{\black} 
\newcommand{\erev}{\black} 

\newcommand{\bblubox}[1]{\begin{tcolorbox}[colframe=blue!75!white,title=#1]}
\newcommand{\eblubox}{\end{tcolorbox}}

\begin{document}

\title{Of Mice and Chickens: \\ Revisiting the RC Time Constant Problem}

\author{Kuni H. Iwasa\\
NIDCD, National Institutes of Health\\
Bethesda, MD 20892, USA}

\date{version 4.1: \today}
\maketitle

\noindent running title: \emph{RC roll-off in hair cells} \\ 
keywords: \emph{auditory frequencies, electrical resonance, piezoelectric resonance, membrane capacitance}

\abstract{Avian hair cells depend on electrical resonance for frequency selectivity. The upper bound of the frequency range is limited by the RC time constant of hair cells because the sharpness of tuning requires that the resonance frequency must be lower than the RC roll-off frequency.  In contrast, tuned mechanical vibration of the inner ear is the basis of frequency selectivity of the mammalian ear. This mechanical vibration is supported by outer hair cells (OHC) with their electromotility (or piezoelectricity), which is driven by the receptor potential. Thus, it is also subjected to the RC time constant problem. Association of OHCs with a system with mechanical resonance leads to piezoelectric resonance. This resonance can nullify the membrane capacitance and solves the RC time constant problem for OHCs. Therefore, avian and mammalian ears solve the same problem in the opposite way.}

\pagebreak

\section*{Introduction}
Auditory frequencies are extremely high for biological cells, which are subjected to their own intrinsic low-pass filter, often referred to as the RC filter. This filter is due to the combination of the membrane capacitance and the membrane conductance. This filter is characterized by the $RC$ time constant, where $R$ is the membrane resistance and $C$ the membrane capacitance. To overcome this low-pass characteristic is a significant challenge for hair cells in the ear. Take mice and chickens, the most commonly used animals for experiments: The auditory range of mice reaches up to 80 kHz \cite{Ehret1976}, while that of chickens extends from 0.25 to 4 kHz \cite{Rebillard1981}. The highest auditory frequency among birds is 9 kHz of barn owls \cite{Konishi1984}. 

The RC time constant problem has been intensely discussed with respect to outer hair cells (OHCs) in the mammalian ear. OHCs are expected to inject power to the oscillation in the inner ear for the sensitivity and frequency specificity of the mammalian ear \cite{bbbr1985,a1987,Dallos2008}. 
However, given the membrane resistance capacitance and membrane resistance of these cells, their RC roll-off frequency is lower than the operating frequency by more than an order of magnitude \cite{ha1992}. This issue, named ``RC time constant problem'' generated various hypotheses  \cite{Dallos1995,Mistrik2009,Spector2003,Rabbitt2009}, including careful re-examination of the membrane resistance \cite{Johnson2011}.

Since all hair cells are subjected to this low pass filter, it is of interest to re-visit this issue, comparing mammalian OHCs with avian short hair cells (SHCs). Biophysical approaches in comparing avian and mammalian hearing would be an interesting addition to more descriptive approaches, such as a parallelism in innervation pattern between avian SHCs and mammalian OHCs \cite{Koeppl2015}. 

The present report shows that reduction of the membrane resistance is an effective strategy for elevating the receptor potential at high frequencies in avian hair cells, which primarily depend on electrical resonance for tuning. A recent summary on vertebrate tuning that depends on this mechanism has been presented by Fettiplace \cite{Fettiplace2020}.

The same strategy does not work for OHCs, which do not have electrical resonance. Instead, OHCs have a significant extra capacitance component, which depends on the membrane potential \cite{s1991} as well as mechanical load \cite{ai1999}. This component is the key to reducing the total membrane capacitance at the operating frequency \cite{Iwasa2017,Iwasa2021}. Therefore, avian ears and mammalian ears employ two contrasting strategies for extending the auditory range to high frequencies: reducing the membrane resistance for the former and reducing the membrane capacitance for the latter.

The scope of the present paper is limited to tuning. Generation of neuronal output is not included because it depends on synaptic transmission, which is a complicated separate process. \brev The results obtained by the present study are largely theoretical predictions, which need to be verified by experiments. \erev

\section*{Avian ears}
Electrical resonance has been demonstrated in turtle hair cells \cite{CrawFett1981}, frog hair cells \cite{HudLew1988} and hair cells in chicks \cite{Fuchs1988,Tan2013} as well as short hair cells \cite{Tan2013} .  The molecular mechanism has been understood as based on the interplay between Ca$^{2+}$ channels and Ca$^{2+}$-activated K$^+$-channels (alias BK channel) \cite{HudLew1988,WuFett1995}. The  resonance frequency is associated with the splice variants of the BK channel in the hair cell \cite{JonesFett1998}.

From a biophysical point of view, the description of electric resonance with Hodgkin-Huxley type equations including those ion channels would be gratifying.  However, such a description involves a good number of parameters \cite{HudLew1988,WuFett1995}. The most significant ones among them are transition rates of the BK channel. For the purpose of finding the frequency limit of electrical tuning, the number of the parameters should be minimized. \brev That can be accomplished by introducing inductance to describe the BK channel \cite{Mauro1970}. That results in \erev the equivalent circuit model proposed by Crawford and Fettiplace \cite{CrawFett1981} is advantageous because the electrical properties of hair cells are quite well described with small number of parameters.

\subsection*{Electric resonance}
\begin{SCfigure}
 \includegraphics[width=0.27\linewidth]{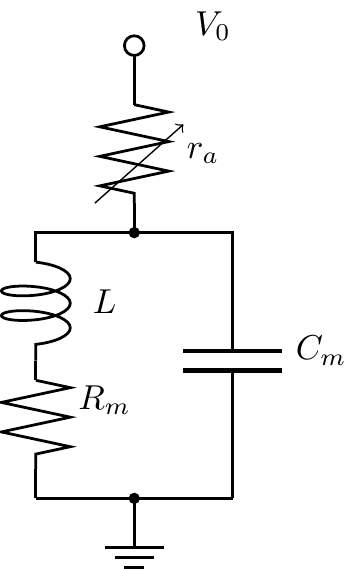}
 \captionsetup{width=1.5\linewidth}
\caption{An equivalent circuit for electric resonance. $r_a$ is hair bundle resistance and $R_m$ is cell body resistance. An interplay between Ca$^{2+}$-channel and Ca$^{2+}$-activated K$^+$ channel \cite{HudLew1988,WuFett1995} leads to electrical resonance, which can be represented by a combination of inductance $L$ and capacitance $C_m$ \cite{CrawFett1981}. }
\label{schematics}
\end{SCfigure}

The equivalent circuit proposed by Crawford and Fettiplace is given in Fig.\ \ref{schematics}. The electrical oscillation, which is realized by an interplay between Ca$^{2+}$-channel and Ca$^{2+}$-activated K$^+$ channel, is introduced by an inductance $L$. As we will see in the following that the resonance frequency depends on the product $C_mL$, where $C_m$ is the membrane capacitance, the inductance $L$ can be interpreted as a parameter, which primarily describes the transition rates of the BK channel.

If the current through the resistor $r_a$  is $I$, then the basic relationships are,
\begin{subequations}
\begin{align}
I&=I_0+I_1,\\
I_0&=\frac{dQ_0}{dt}, \qquad Q_0=CV \\
V&=\left(R +L\frac {d}{dt}\right)I_0
\end{align}
\end{subequations}
These relationships lead to
\begin{align}
LC\frac{d^2V}{dt^2}+RC\frac{dV}{dt}+V=RI+L\frac{dI}{dt}.
\end{align}

Now input sinusoidal current $I=\langle I\rangle+i_\mathrm{in}\exp[i\omega t]$, and let $V=\langle V\rangle v\exp[i\omega t]$. Then the differential equation turns into
\begin{align}\label{eq:omega_eq1}
(-\overline\omega^2+i\overline\omega/\overline\omega_{rc}+1)v=(R+i\omega L)i_\mathrm{in}
\end{align}
where
\begin{subequations}
\label{eq:f_ratio}
\begin{align} \label{eq:omega_r}
\overline\omega&=\omega/\omega_r, \qquad \omega_r^2=1/(LC), \\ \label{eq:omega_eta}
\overline\omega_{rc}&=\omega_{rc}/\omega_r, \qquad \omega_{rc}=1/(RC).
\end{align}
\end{subequations}
A larger value for the RC roll-off frequency $\overline\omega_{rc}$ gives rise to a sharper resonance peak. For this reason, it is known as the quality factor $Q$.
Eqs.\ \ref{eq:f_ratio} indicates that the ratio $L/R$ can be expressed by $\omega_\eta/ \omega_r^2$. Thus, Eq.\ \ref{eq:omega_eq1} can be expressed by
\begin{align}\label{eq:omega_eq}
(-\overline\omega^2+i\overline\omega/\overline\omega_{rc}+1)v=R(1+i\overline\omega \cdot \overline\omega_{rc} )i_\mathrm{in}.
\end{align}

\begin{figure}[h]
\hspace{0.05\linewidth} A \hspace{0.3\linewidth} B  \hspace{0.3\linewidth} C
\vspace{-5mm}
\begin{center}
\includegraphics[width=0.33\linewidth]{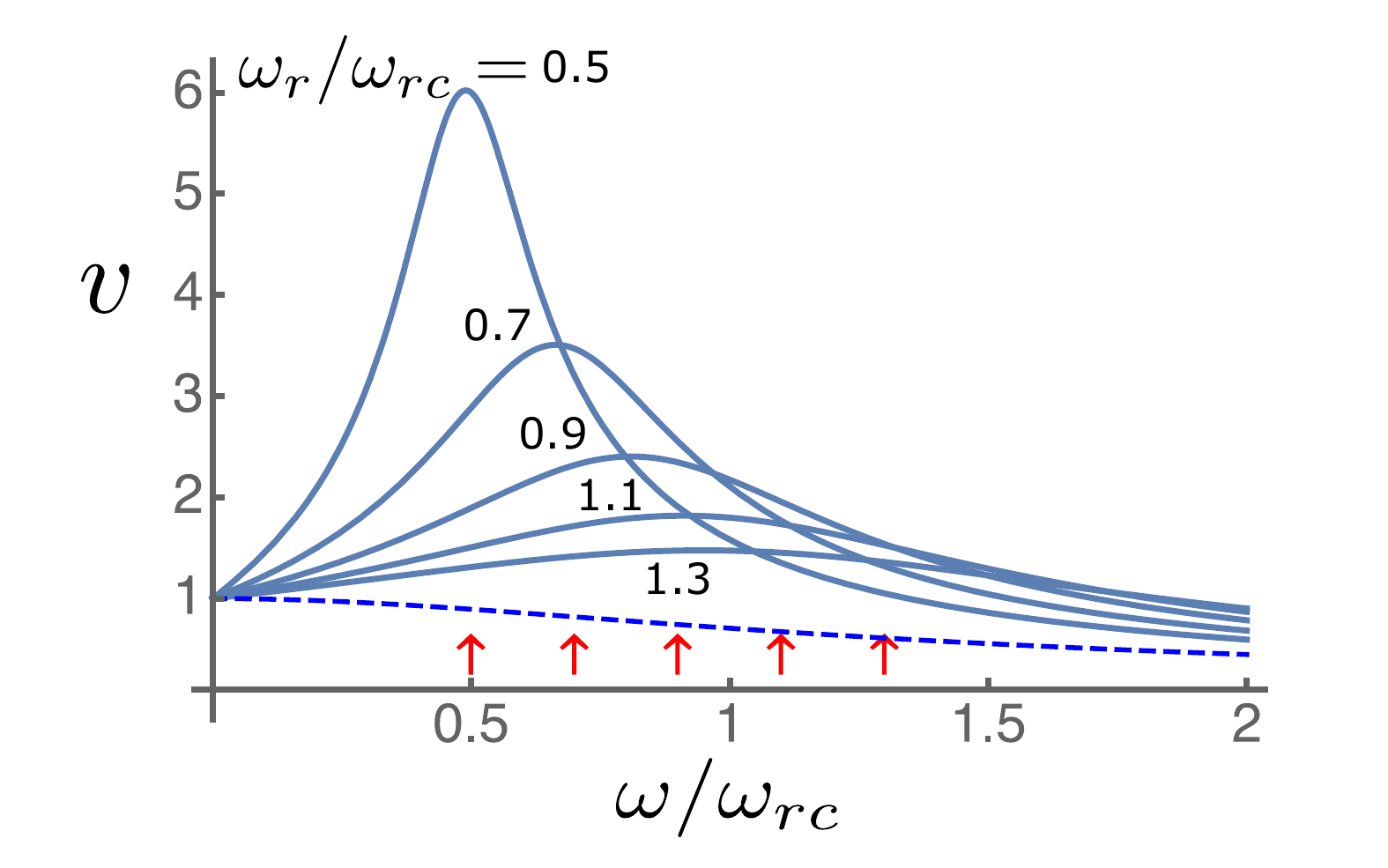}\ \
\includegraphics[width=0.32\linewidth]{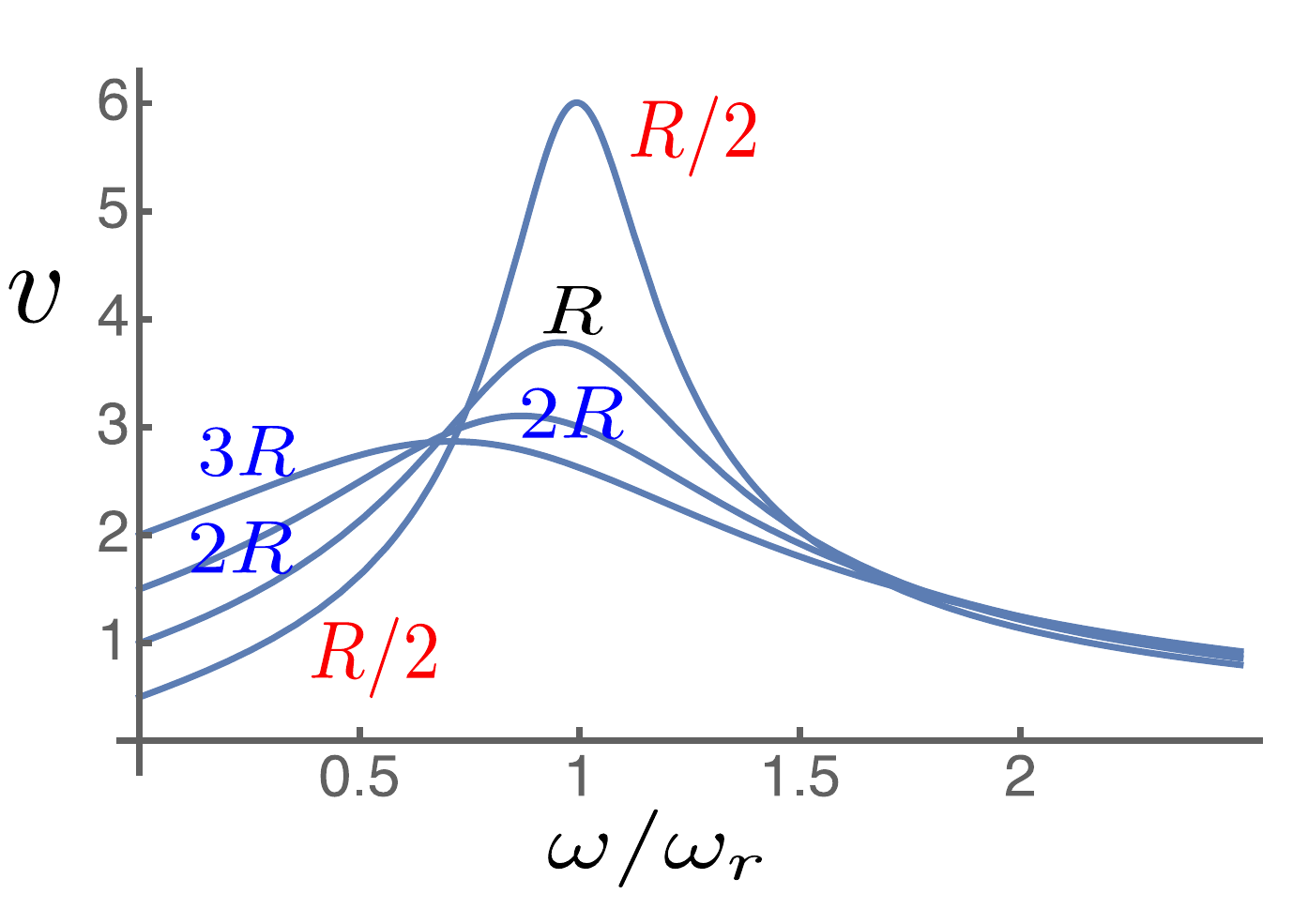}\ \
\includegraphics[width=0.3\linewidth]{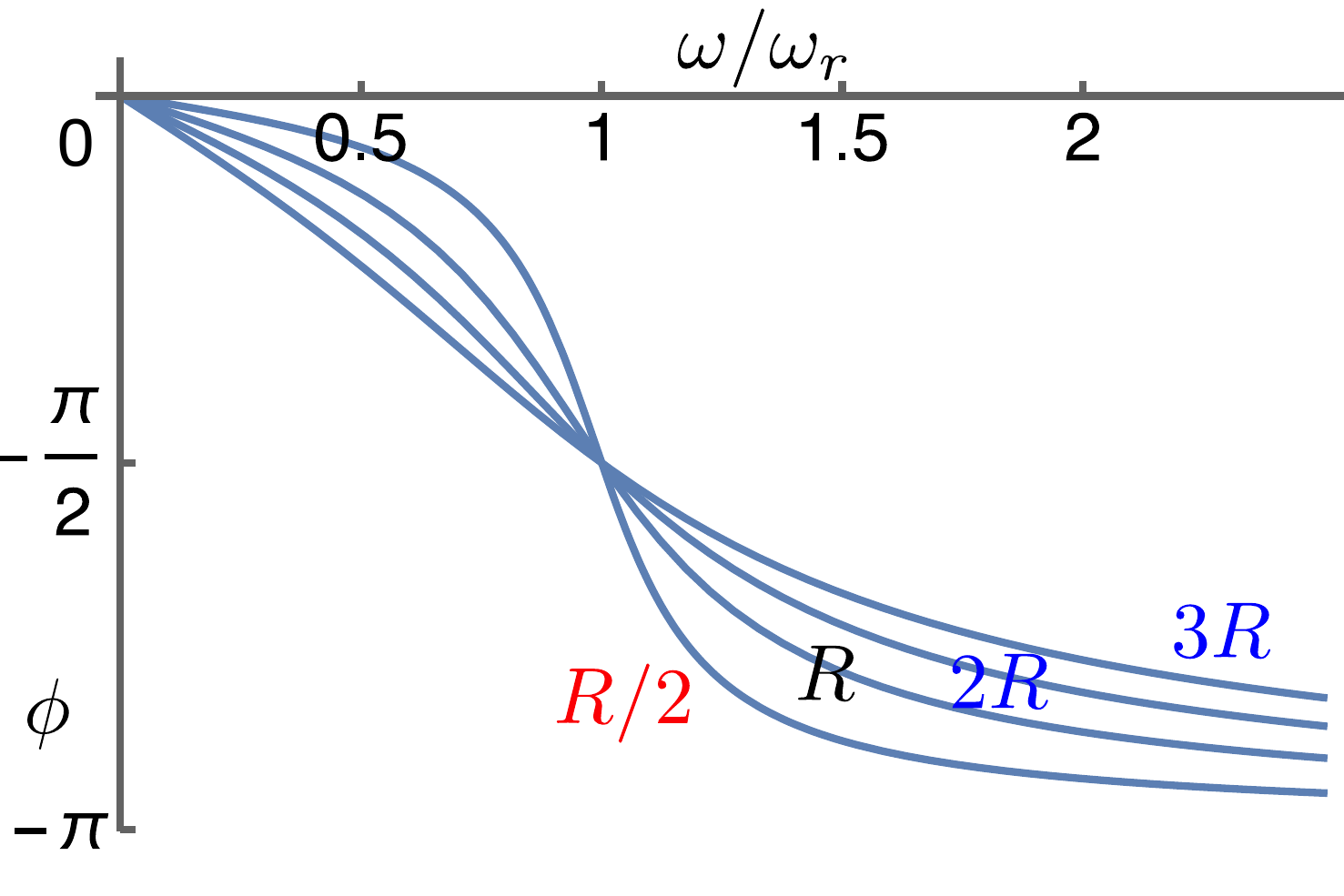}
\caption{The effect of resonance frequency and RC roll-off frequency on the amplitude $v$ of the receptor potential (Eq.\ \ref{eq:r_ratio}).  The amplitude $v$ is normalized at $\omega=0$. A:  The effect of changing the resonance the ratio $\omega_{rc}$. The frequency axis is normalized by the RC roll-off frequency $\omega_{rc}$. The values for $\omega_r/\omega_{rc}$ are 0.5, 0.7, 0.9, 1.1, and 1.3 (red arrows). The broken line shows the response without resonance. The value of $R$ is set to unity. B: The effect of the membrane resistance.  \brev Notice the peak shift to lower frequencies as well as the broadening as the resistance increases. The frequency axis is normalized by the resonance frequency. Relative values of the membrane resistance are indicated. Here ``$R$'' represents the condition $\omega_{rc}/\omega_r=1.5$. C: The phase response that corresponds to B. The phase is approximately constant near $\omega/\omega_r=1$. \erev} 
\label{fig:resistance}
\end{center}
\end{figure}

The resulting tuning curve is shown in Fig.\ \ref{fig:resistance}. A frequency dependent term in the nominator somewhat affects the outcome but the difference is not very large. The existence of a peak frequency requires the condition $\overline\omega_{rc} >1$.

Now, current input $I$ is due to changes in the hair bundle resistance $r_a$, expressed by
\begin{subequations}
\begin{align}
I&=(V_0-V)/r_a \quad \mathrm{with} \\
 r_a&=\langle r_a \rangle(1-r\exp[i\omega t]).
\end{align}
\end{subequations}
If the relative change in the hair bundle resistance $r$ is small, we obtain
\begin{align*}
I \approx \langle I\rangle+\left(\langle I\rangle r-\frac{v}{\langle r_a\rangle}\right)\exp[i\omega t],
\end{align*}
where $\langle I \rangle=(V_0-\langle V\rangle)/\langle r_a\rangle$. Thus, we obtain
\begin{align}
i_\mathrm{in}=\langle I\rangle r-\frac{v}{\langle r_a\rangle}
\end{align}
Thus Eq.\ \ref{eq:omega_eq} is replaced by
\begin{align}\label{eq:r_ratio}
\left(-\overline\omega^2+i\overline\omega\left(\frac{1}{\overline\omega_{rc}}+\overline\omega_{rc}\frac{R}{\langle r_a\rangle}\right)+1+\frac{R}{\langle r_a\rangle} \right)v=\langle I \rangle R(1+i\overline\omega\cdot\overline\omega_{rc})r.
\end{align}
The frequency dependence approaches the case of sinusoidal input current (Eq.\ \ref{eq:omega_eq}) if $R/ \langle r_a\rangle \rightarrow 0$, as would be expected. That is indeed the physiological condition.

The above equations show that the efficacy of electric tuning requires that $\omega_\eta$, the roll-off frequency of the RC filter, must be higher than the resonance frequency $\omega_r$. In other words, the  roll-off frequency of the intrinsic RC filter dictates the frequency bandwidth. 

\subsection*{Control by membrane resistance}
The membrane capacitance depends on the surface area. Tall hair cells (THCs) on the neural side of the chicken basilar papilla do not show much difference in apparent membrane area \cite{Gleich2000} and indeed in the membrane capacitance \cite{Fuchs1988,Tan2013}. \brev While SHCs appear to show smaller area toward the base, the membrane capacitance does not show such a trend, likely attributable to the larger surface area of the hair bundle in basal cells  \cite{Tan2013}.  \erev

A reduction of the membrane resistance to enhance the receptor potential may appear counterintuitive because that reduces the receptor potential in the low frequency range (Fig.\ \ref{fig:resistance}B). However, this effect can sharply enhance the  receptor potential near the resonance frequency because it elevates the RC roll-off frequency (Fig.\ \ref{fig:resistance}B). The net gain near the resonance frequency more than compensates for the reduction due to the decreased $R_m$.

This strategy of reducing the membrane resistance $R_m$ is, unlike reducing the membrane capacitance $C_m$, not physically constrained. However, it could be metabolically taxing. To achieve an RC roll-off frequency of \brev 5 kHz, $R_m$ must be $\sim$4 M$\Omega$. \erev In addition, electrical resonance itself requires higher metabolic rates for high frequencies because the resonance mechanism requires oscillation of intracellular Ca$^{2+}$ ion concentration. Oscillation of intracellular Ca$^{2+}$ ion concentration during each cycle requires an increased power consumption for calcium extrusion to operate at higher frequencies. 

Since the resonance frequency must be lower than the RC roll-off frequency,  even with a roll-off frequency of \brev 5 kHz, \erev the reasonably sharp resonance must be\brev  $(5/Q)$ \erev kHz, where $Q$ is the value of the quality factor. To achieve quality factor 5, the value common for turtle hair cells \cite{CrawFett1981}, the resonance frequency must be \brev 1 kHz. \erev

\brev An experimental study on THCs from E18 -- E19 birds shows resonance frequency between the fractional distance from apex of 0.1 to 0.6 at 23$^\circ$ and 33$^\circ$C \cite{Tan2013}.  An extrapolation the body temperature of 40$^\circ$C appears to be $\sim$1 kHz at the fractional distance from apex of 0.7 \cite{Tan2013,Fettiplace2020}. 

\erev For this reason, achieving the upper limit of 4 kHz for the chicken \cite{Rebillard1981} may require an additional mechanism, which improves the quality factor.

\subsection*{Motion of tectorial membrane}

\begin{SCfigure}
\includegraphics[width=0.35\textwidth]{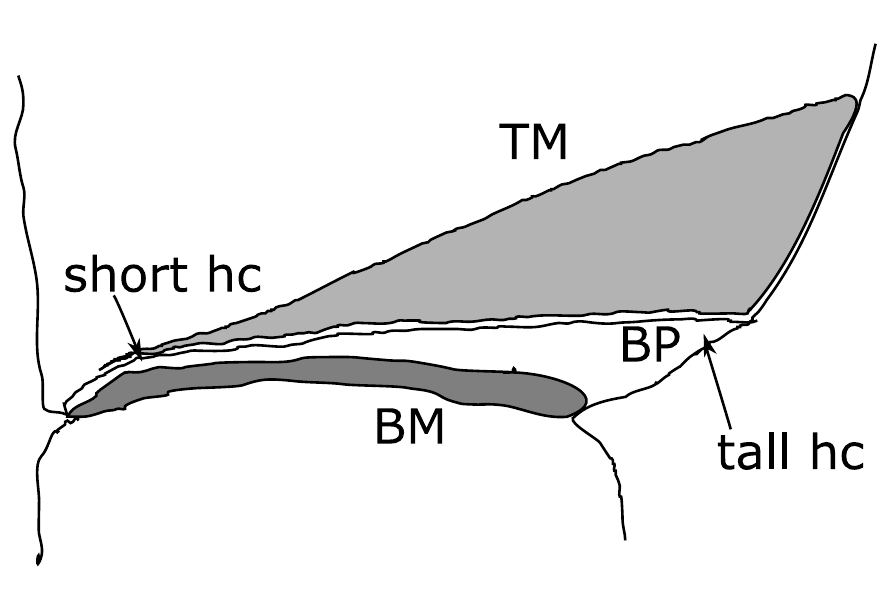}
\captionsetup{width=1.0\linewidth}
  \caption{\small{The cross-section of the avian inner ear. TM: the tectorial membrane, BM: the basilar membrane, BP: the basilar papilla. After refs.\  \cite{Gleich2000,Tan2013}.}}
    \label{fig:anatomy}
\end{SCfigure}

A mechanism that can enhance high-frequency hearing could be based on the tonotopic organization of the avian ear. One such mechanism could be the movement of the tectorial membrane (TM), which can enhance hair bundle input\brev \cite{Steele1997,Fettiplace2020}. \erev The avian TM, which is softer than the mammalian one, is tapered toward the abneural side (Fig. \ref{fig:anatomy}) and the tapering increases toward the base. \brev That is a distinctive feature of the avian TM. The mammalian TM does not show such tapering or a lateral shift in morphology. \erev

Like OHCs in the mammalian ear, SHCs are located nearly above the middle of the basilar membrane, the movement of which in the up-down direction \cite{Gummer1987} would effectively transmit bending stress to the SHCs' hair bundles. In contrast, THCs, which function as the sensor, are located away from the center of the basilar membrane, at a location seemingly unsuitable for sensing the motion of the basilar membrane directly. Thus it is likely SHCs are involved in amplification by an active process of the hair bundle, transmitting the mechanical vibration of the basilar membrane to the hair bundles of THCs. 

\brev Specifically for the avian ear, an active process in the hair bundle is associated with the receptor potential of SHCs \cite{Beurg2013}, rather than more generic active process of the hair bundle \cite{Steele1997,cho-hud1998,Tinevez2007,Sul2009c}. It is also prestin dependent \cite{Beurg2013}.

\erev In the following section, we show a simple description of a vibrational mode that can selectively enhance hair bundle input.
Since hair cell bundles are stimulated by bending, it would be logical to assume that shear motion between the basilar papilla and the TM stimulates the hair bundles.  Although the orientation of hair bundles forms domain structure in a more apical region, it is uniform in the basal region \cite{Koeppl1998,gle-man2000}. 

\begin{SCfigure}
  \includegraphics[width=0.35\textwidth]{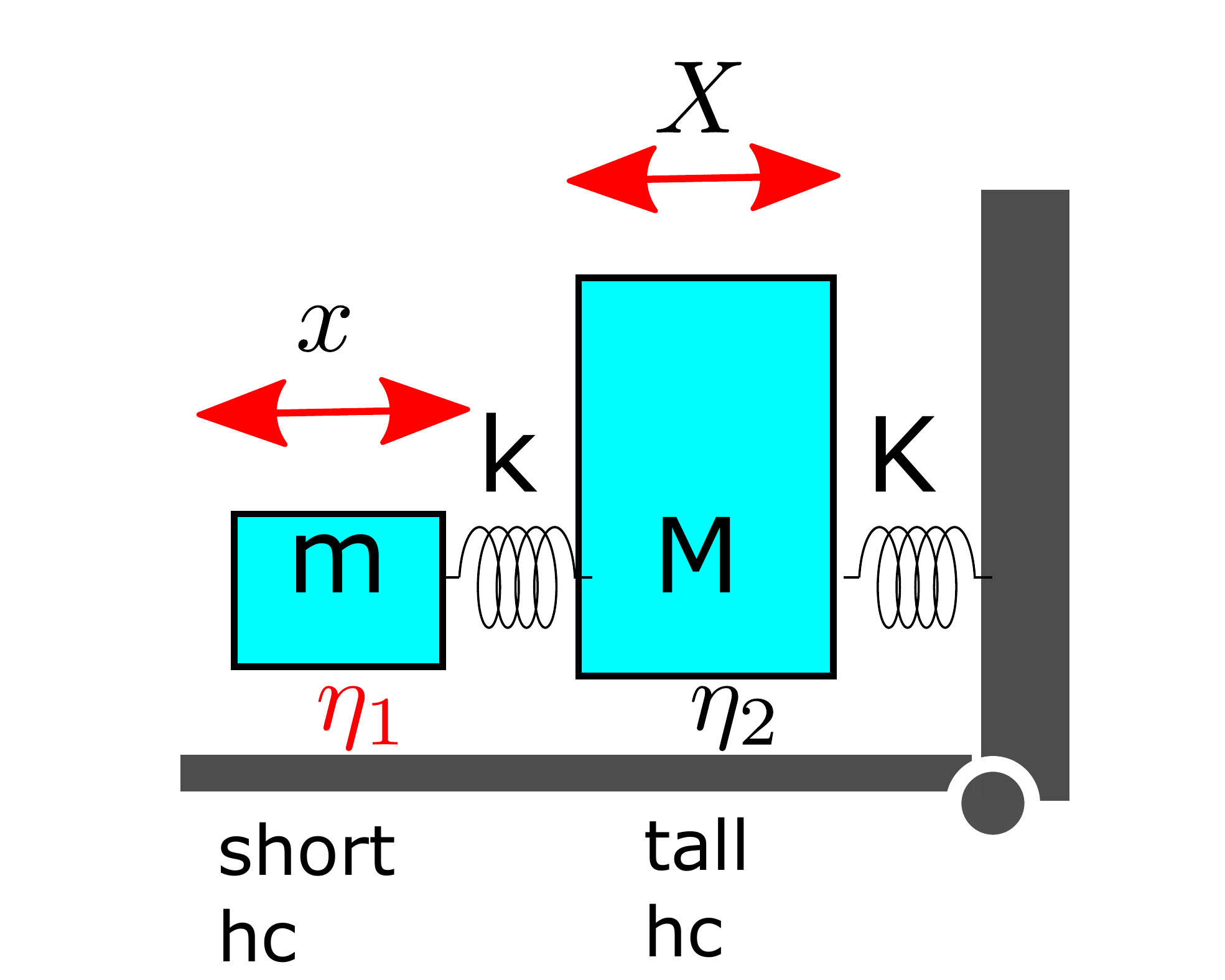}
\captionsetup{width=1.2\linewidth}
  \caption{\small{A possible mode of motion of the avian inner ear. Two squares represent two parts of the tectorial membrane. The smaller one on top of SHCs and the larger one on top of THCs. The tapering the tectorial membrane is represented by the mass ratio $M/m$.}}
    \label{fig:modes}
\end{SCfigure}

\brev For simplicity, specifically to reduce the number of parameters, let us make the following main assumptions. 
\begin{enumerate}
\item Hair bundles are aligned perpendicular to the lateral axis, consistent with the morphological observations on the basal area.
\item The tapering of the TM is represented by a two-mass model (Fig.\ \ref{fig:modes}). A smaller mass $m$ is located at the top of SHCs and another with a larger mass $M$ is located at the top of THCs \cite{iwasa2014}.
\item Hair bundles under mass $M$ are passive, and hair bundles under mass $m$ are active.
\item The active process of the hair bundles under mass $m$ is represented by negative drag.
\item Lateral connectivity of the TM can be ignored.
\end{enumerate}

\erev

More quantitatively, the small mass $m$ is connected to the larger mass $M$ with a spring with stiffness $k$ and the larger mass $M$ is connected with the wall by another spring with stiffness $K$. Let $X$ be the displacement of the bigger mass from its equilibrium position, and that of smaller one be $x$. Let $\eta_2$ be viscous drag on mass $M$ and $\eta_1$ be negative drag on mass $m$ to mimic the amplifying role of SHCs (Fig.\ \ref{fig:modes} A). 

The set of equations is given by,
\begin{subequations}
\label{eq:motion}
\begin{align} \label{eq:motion_m}
\left(m\frac{d^2}{dt^2}-\eta_1\frac{d}{dt}\right)x-k(X-x)&=f_0\exp[i\omega t] \\
\label{eq:motion_M}
\left(M\frac{d^2}{dt^2}+\eta_2\frac{d}{dt}+K\right)X-k(X-x)&=0,
\end{align}
\end{subequations}

where $f_0\exp[i\omega t]$ is a periodic external force of angular frequency $\omega$ applied to mass $m$, which is located just above the center of the basilar membrane. Here it is assumed that up-down motion of the basilar membrane results in this external force. Following Steele \cite{Steele1997}, let $\omega_0$ be the resonance frequency of the tectorial membrane, i.e. $\omega_0^2=K/M=k/m$ and seek a steady state solution. 

\begin{figure}
A \hspace{0.5\linewidth} B\\
 \includegraphics[width=0.5\textwidth]{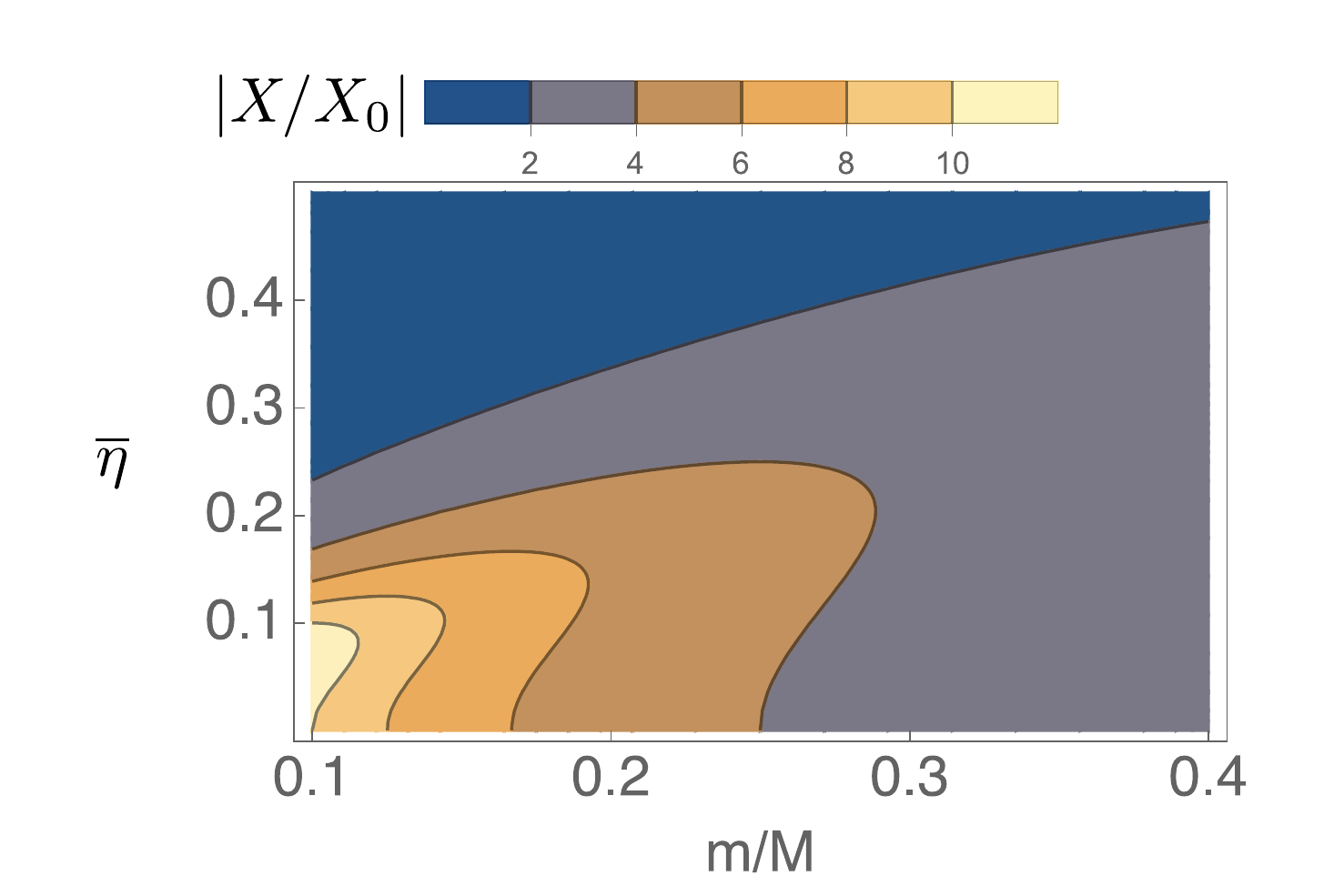} \ \
  \includegraphics[width=0.45\textwidth]{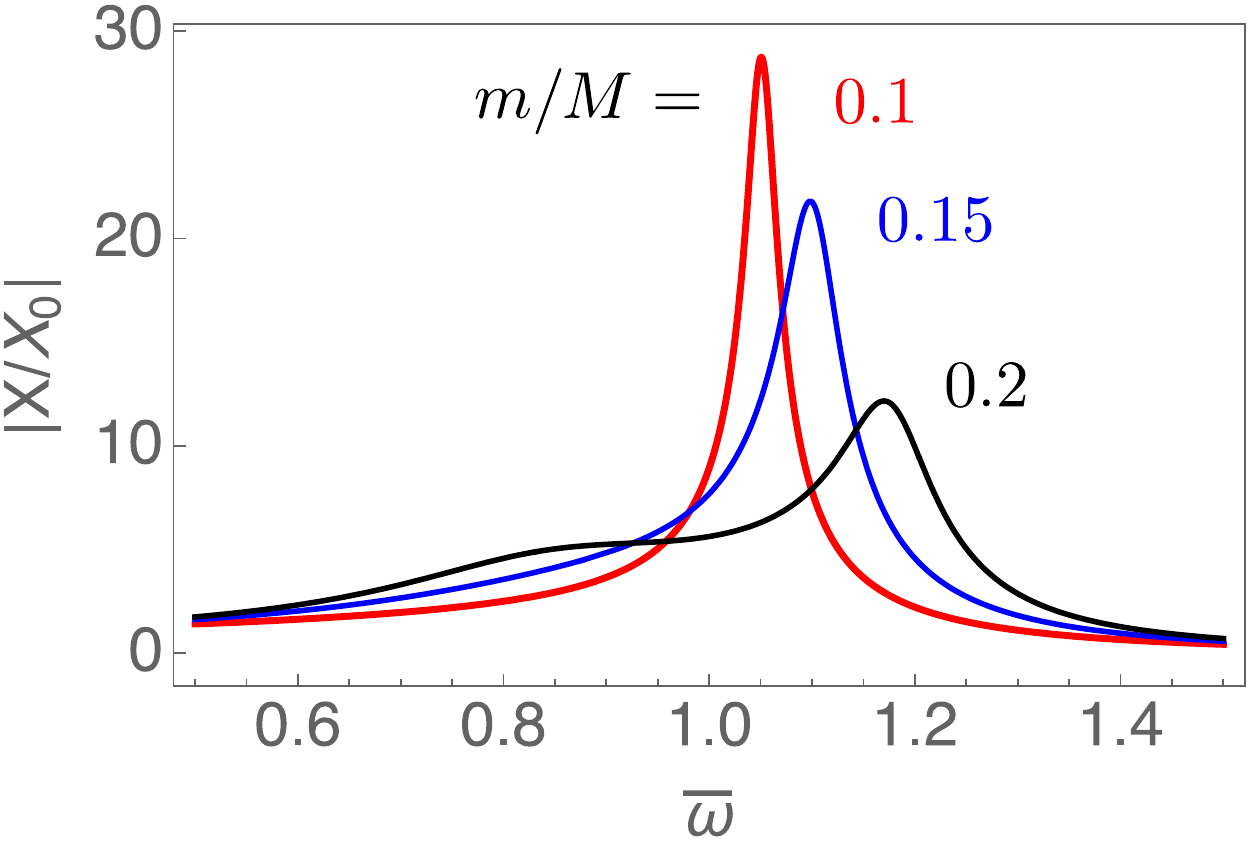} 
  \caption{\small{\brev The effect of the mass ratio. A: Normalized displacement $|X/X_0|$ of mass $M$ at the resonance frequency $\omega_0$, is plotted against the mass ratio $m/M$ and the reduced drag coefficient $\overline{\eta}=\eta/(M\omega_0)$. The normalization factor is $X_0=f_0/K$.  B: The frequency dependence at $\overline{\eta}=0.11$. Traces are for $m/M=0.1$, 0.15, and 0.2 from the left. The abscissa is the normalized frequency $\omega/\omega_0$. The half-peak frequency $\Delta\overline\omega$ for the peak of $m/M=0.1$ is 0.05, indicating the quality factor of 20. \erev
}}
\label{fig:mass_ratio}
\end{figure}

\brev For a special case of $\eta_1=\eta_2(=\eta)$, which reduces the number of parameters, the force applied to the mass $M$ \brev at the mechanical resonance frequency \erev is contour-plotted against the mass ratio $m/M$  and the reduced drag $\overline{\eta}=\eta/(M\omega_0)$(Fig.\ \ref{fig:mass_ratio}A). For a given value of $M$, this force is larger for smaller values of the reduced drag $\overline{\eta}$ where the ratio $m/M$ is smaller. The plot shows that the force at the mechanical resonance frequency $\omega_0$ is larger for large ratio $M/m$. This result is consistent with the morphological observation that the sharpness of tapering of the TM increases from the apical to the basal end \cite{Smith1985,Koeppl1998}. One can argue that the force relevant to the transduction of THCs would be better related to drag $\eta dX/dt$ rather than the elastic force $KX_0$, where $X_0$ is the amplitude of $X$ at frequency $\omega_0$. That does not change the optimal mass ratio because drag is represented by $\overline{\eta}KX_0$, maximizing at the same mass ratio. 

\brev To evaluate the improvement of the quality factor, the force applied to the mass $M$ is plotted against the frequency (Fig.\ \ref{fig:mass_ratio}B). Resonance peaks are quite sharp, and are capable of achieving the quality factor of 20 near the peak of electrical resonance frequency. 

It is likely that such a large value for the quality factor is an overestimate due to the simplifying assumptions. For example, the lateral connectivity of the tectorial membrane and the two-oscillator model, which introduces active process as negative drag are likely oversimplifications. 

\begin{figure}
A \hspace{0.5\linewidth} B\\
 \includegraphics[width=0.5\textwidth]{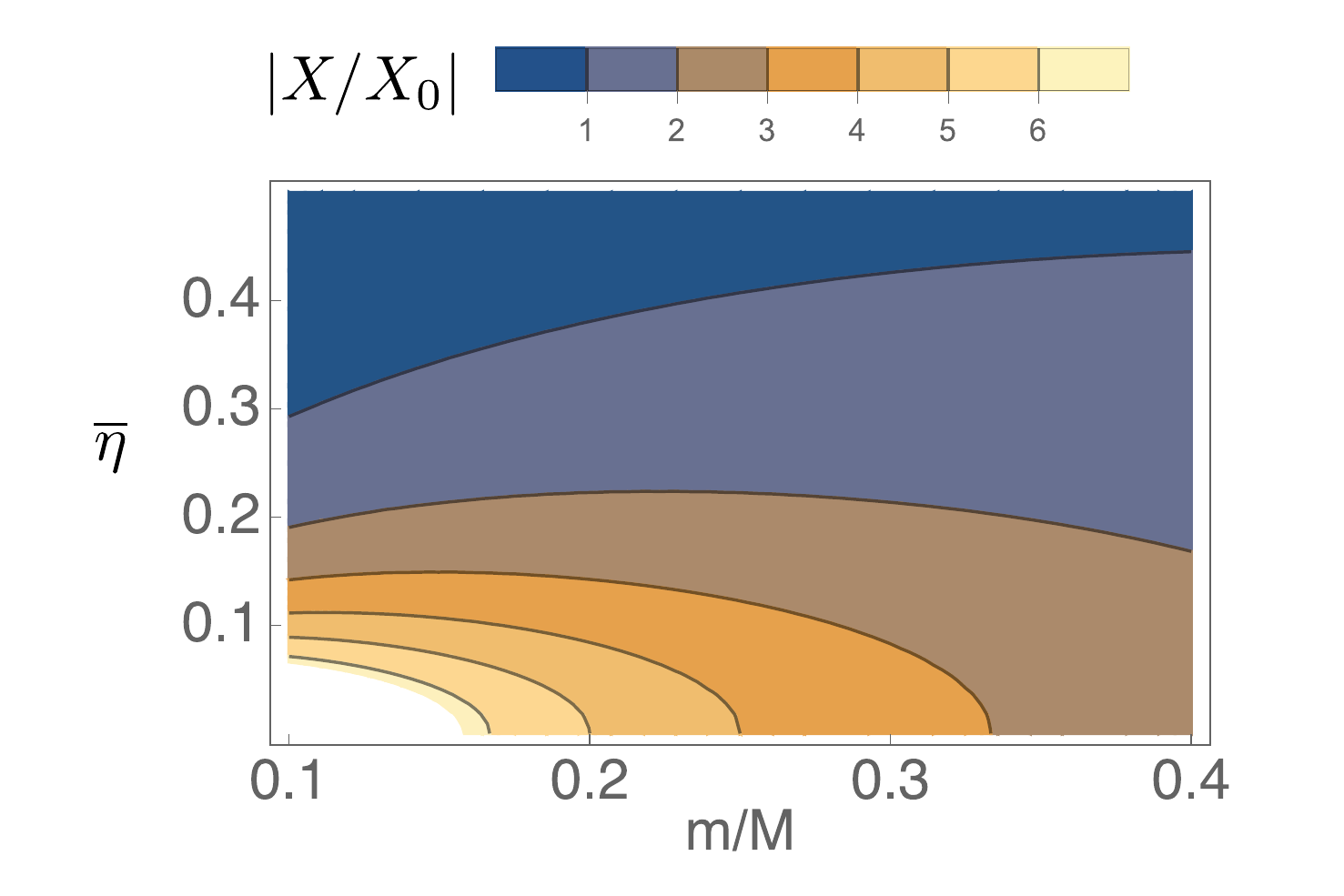} \ \
 \includegraphics[width=0.45\textwidth]{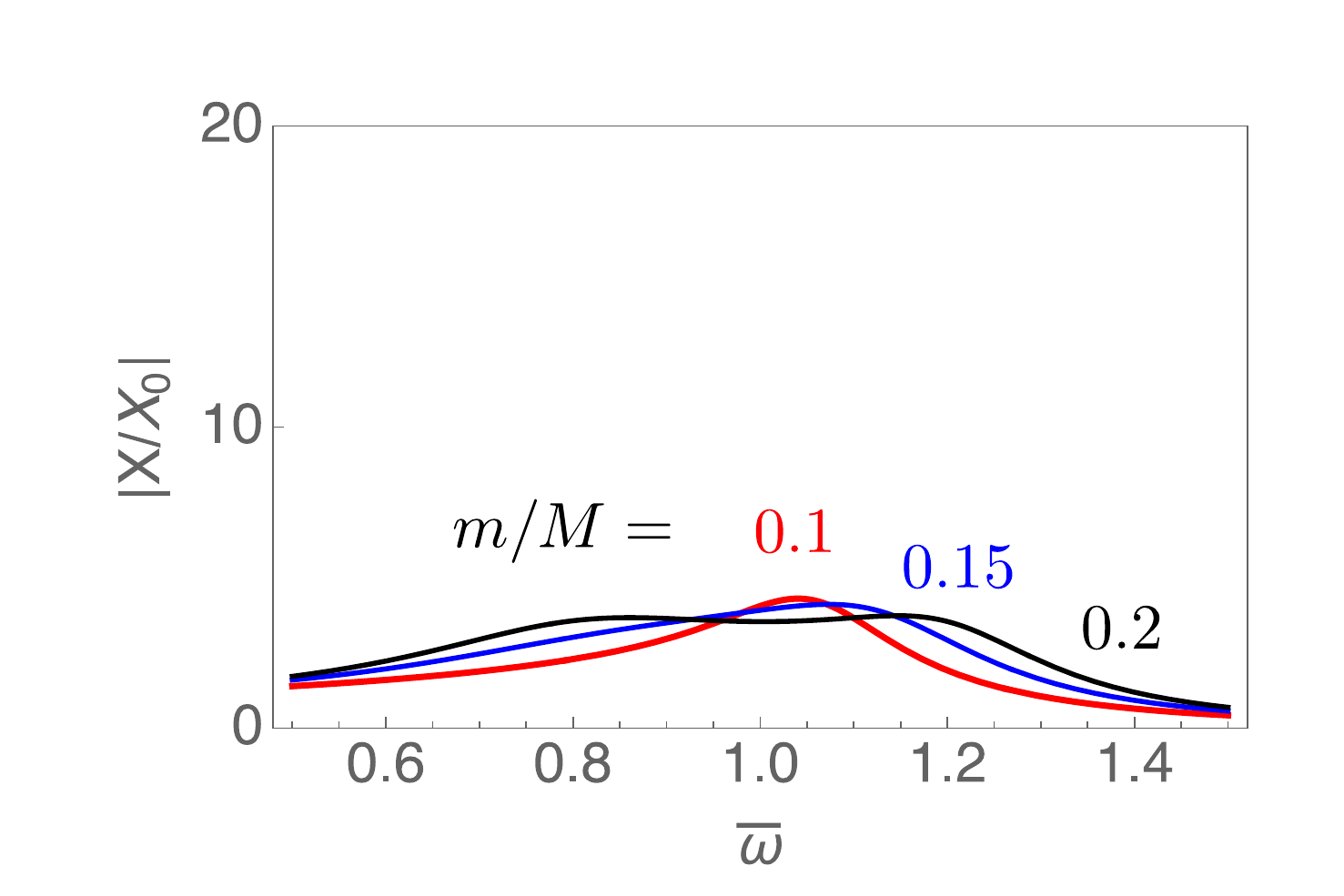} 
 \caption{\small{ \brev The effect changing anti-drag to drag. The sign of $\eta_1$ in Eq.\ \ref{eq:motion_m} is changed from minus to plus. A: Normalized displacement $|X/X_0|$ of mass $M$ at the resonance frequency $\omega_0$, is plotted against the mass ratio $m/M$ and the reduced drag coefficient $\overline{\eta}=\eta/(M\omega_0)$. The normalization factor is $X_0=f_0/K$.  B: The frequency dependence at $\overline{\eta}=0.11$. Traces are for $m/M=0.1$, 0.15, and 0.2 from the left. The abscissa is the normalized frequency $\omega/\omega_0$. \erev
  }}
\label{fig:mass_ratio2}
\end{figure}

The assumption of negative drag may need some clarification. The first question is how to describe the condition, where the active process in the hair bundle of SHCs, which creates negative drag, is turned off. It would be natural to assume the negative drag under the mass $m$  is replaced by simple (positive) drag, similar to that under the larger mass $M$. Under this condition, frequency selectivity largely disappears (Fig.\ \ref{fig:mass_ratio2}).

This issue leads to the second question on the magnitude of the negative drag. While drag increases with increasing frequency, the active force generated in hair bundles may not increase with frequency. The experimental observation that the hair bundles of SHCs are driven by the receptor potential via avian prestin \cite{Beurg2013} indicates that this active force depends on the magnitude and the phase of the receptor potential of SHCs. 
Since the receptor potential of SHCs is subjected to the RC filter, its magnitude decreases with increasing operating frequency. Under such conditions, it is hard to justify negative drag as expressed by Eqs.\ \ref{eq:motion} as an approximation. Thus, the validity of this approximation deteriorates at frequencies beyond $\omega_{rc}$. Hence $\omega_0 \lesssim \omega_{rc}$.

Regarding the phase, we could assume Eqs.\ \ref{eq:motion} are satisfied at the phase of the resonance at the frequency $\omega_r$, which is not significantly shifted by the membrane resistance (See Fig.\ \ref{fig:resistance}C). Then, the validity of Eqs.\ \ref{eq:motion} is limited near the electrical resonance frequency $\omega_r$.

With all these reservations, we could still expect that the resonance of the tectorial membrane improves the quality factor with the active process in the hair bundles of SHCs, even though the frequency range of the ear is still limited by the low pass nature of the RC filter in SHCs.

\erev Motion of the tectorial membrane, however, has not been detected so far \cite{XiaOgh2016}. \brev Since THCs, not being located above the basilar membrane, cannot be stimulated without moving the tectorial membrane, the reason must be sought in the technical difficulties in the experiment. The major reason could be \erev the low optical density of the avian tectorial membrane, which lacks collagen. In addition, the border between the tectorial membrane and the basilar papilla could be optically blurry. Unlike the flat surface of the mammalian tectorial membrane, the avian TM has an elaborate structure on the side facing the basilar papilla. A thin veil-like structure descends to the basilar papilla, surrounding each hair cell \cite{Goodyear1994}. A dome like recess above each hair cell and each hair bundle makes contact with the tectorial membrane at the dome \cite{Goodyear1994}. 

\section*{Mammalian ears}
Of mature mammalian cochlear hair cells, neither inner hair cells (IHCs), nor OHCs (OHCs) show spontaneous firing or electrical oscillation. The organ of Corti, into which hair cells are incorporated, shows marked displacements, which are associated with the traveling wave on the basilar membrane, during acoustic stimulation of the ear \cite{bekesy1952}.

Outer hair cell motility has been thought to provide power into the motion in the cochlea to amplify the oscillation in the organ of Corti \cite{bbbr1985,a1987}. However, the frequency dependence of such a function had been a puzzle, often referred to as ``RC time constant problem,'' which many investigators tried to solve \cite{Dallos1995,Mistrik2009,Spector2003,Rabbitt2009,Johnson2011,OMaoileidigh2013,Iwasa2016,Iwasa2017,Liu2017,Rabbitt2020}. The reason was that the roll-off frequencies of OHCs are much lower than their operating frequencies, suggesting a significant attenuation of the effectiveness. 

 \subsection*{Two components of the capacitance} 

A puzzle in evaluating the RC time constant is the value of the membrane capacitance to use because OHCs have two major components of their membrane capacitance. One of the components has been referred to as the ``regular capacitance.'' It does not depend on the membrane potential, quite similar to the membrane capacitance of most cells. Indeed, it is consistent with the standard specific capacitance of 1$\mu$F/cm$^2$ \cite{cole}. The other main component is often called ``nonlinear capacitance'' because it shows a bell-shaped dependence on the membrane potential \cite{s1991,Santos-Sacchi1998a}. It is associated with prestin \cite{zshlmd2000}, a member of the SLC family of membrane proteins, which confers voltage-dependent motility on OHCs \cite{a1987}, often referred to as electromotility.
For the purpose of clarifying the origins, let us call the former the ``structural'' capacitance, and the latter ``prestin'' capacitance. 

Prestin capacitance cannot be ignored because at its peak voltage it is larger than the structural capacitance in basal OHCs. It shows frequency dependence \cite{ga1997} and sensitivity to mechanical force and constraints \cite{i1993,ga1994,ai1999}, consistent with mechanoelectric coupling, or piezoelectricity, of this molecule \cite{doi2002}.

\subsection*{Piezoelectric resonance}
OHCs undergo reduction of their length in response to depolarization and elongation in response to hyperpolarization.  Inversely, externally imposed length changes of OHCs induce membrane current, consistent with piezoelecticity \cite{doi2002}.

\begin{figure}
\includegraphics[width=0.25\linewidth]{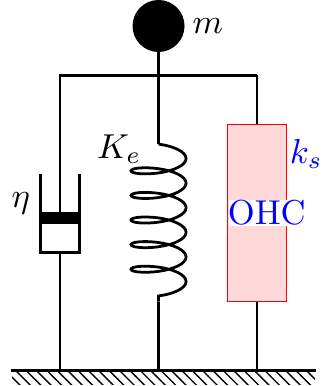}\ \ 
\begin{minipage}[b]{0.5\linewidth}
\caption{
\small{OHCs incorporated into a mechanically resonating system. $K_e$ is the stiffness of the elastic load, $\eta$ the drag coefficient, and $m$ the inertial mass. The intrinsic stiffness of the OHC is $k_s$. }
\label{fig:mech_schem}
}
\end{minipage}
\end{figure}

For this reason, if OHCs are associated with a mechanically resonating system, piezoelectric resonance, \brev a mechanism that can overcome the RC time constant, is expected \cite{mh1994,Weitzel2003}. \erev At frequencies lower than the resonance point, the electric charge in prestin moves in phase with the rest of the capacitor charge, which is 90 degrees in advance of the voltage waveform.  Near the mechanical resonance frequency of the system, the mechanical oscillation together with the length changes of the OHC is delayed by 90 degrees from the voltage waveform. Under this condition, the current carried by prestin goes opposite to the capacitor current, reducing or even reversing the total current, resulting in reduced or negative capacitance, depending on the relative significance of the charge of the motile element. This is equivalent to piezoelectric resonance and the motile membrane protein pumps electrical energy into mechanical vibration without being hindered by the RC time constant problem, as described below.

\subsubsection*{A model system}

Let an OHC be incorporated into a system with mechanical resonance as illustrated in Fig.\ \ref{fig:mech_schem}. The description largely follows the one-dimensional model \cite{Iwasa2017}, rather than the more realistic cylindrical model \cite{Iwasa2021} for simplicity.
 The focus here is effect of resonance on the membrane capacitance.

\subsubsection*{Motile element} 
Let the motile element in the lateral membrane of the OHC have two states, the long state $L$ and the short state $S$, and let $P$ be the fraction of the motile element in state $L$. The natural length of the cell would be $X_0+anP$, where $a$ is the contribution of a single motile element to the cell length accompanied by conformational change from $S$ to $L$. The same conformational change transfers a charge $q$ across the membrane. The number of the motile elements in the cell is $n$.

With the elastic load $K$, the displacement $X$ due to the motile unit is given by
\begin{align}
 X=anP\;k_s/(k_s+K_e),
\end{align}
where $k$ is the structural stiffness of the cell.
 In addition, assume that $P_\infty$ at equilibrium follows the Boltzmann distribution
\begin{align}
 P_\infty/(1-P_\infty)=\exp[-\beta(G_0+qV+aF_o)]
\end{align}
where $\beta=1/(k_BT)$ with Boltzmann's constant $k_B$ and the temperature $T$. The constant term is represented by $G_0$. The quantities $V$ and $F_o$ are, respectively, the membrane potential and force applied from the external elastic load, which is associated with axial displacement $X$ of the cell.  Since the stiffness of the external elastic load is $K$, $F_o=KX$.

\subsubsection*{Equation of motion} 
The equation of motion of the system that is illustrated in Fig.\ \ref{fig:mech_schem} can be expressed by
\begin{align}
 m \; d^2X/dt^2+\eta dX/dt=k_s(X_\infty-X),
 \label{eq:x_eqn}
\end{align}
where $X_\infty=anP_\infty\; k_s/(k_s+K_e)$ is the displacement that corresponds to equilibrium, $m$ is the mass, and $\eta$ drag coefficient. The inertia term can be justified if the system is not far from equilibrium \cite{Iwasa2021}.
The difference between the current displacement $X$ and equilibrium displacement $X_\infty$ is the driving force. The equation of motion (\ref{eq:x_eqn}) can be expresses using variable $P$
\begin{align}
m \; d^2P/dt^2+\eta dP/dt=(k_s+K_e)(P_\infty-P).
 \label{eq:p_eqn}
\end{align}

\subsubsection*{Prestin capacitance}
Now consider the response to a sinusoidal voltage waveform of small amplitude $v$ and angular frequency $\omega$. Let $p$ be the corresponding small amplitude of $P$.
\begin{align*}
V=V_0+v\exp[i\omega t], \quad
P=P_0+p\exp[i\omega t].
\end{align*}
Then the equation of motion (\ref{eq:p_eqn}) is transformed into
\begin{align}
 p=\frac{-\gamma q}{-\overline\omega^2+i\overline\omega/\overline\omega_\eta+\alpha^2}\;v,
\label{eq:p}
\end{align}
where $\overline\omega(=\omega/\omega_r)$ is the frequency normalized to the mechanical resonance frequency $\omega_r=\sqrt{(k_s+K_e)/m}$, $\overline\omega_\eta$ is normalized viscoelastic roll-off frequency, $\gamma=\beta P_0(1-P_0)$, and $\alpha^2=1+\gamma a^2nk_sK_e/(k_s+K_e)$. The coefficients $\gamma$ and $\alpha$ originate from the expansion of the exponential term of $P_\infty$.

\begin{figure}
A \hspace{0.5\linewidth} B \\
\includegraphics[width=0.48\linewidth]{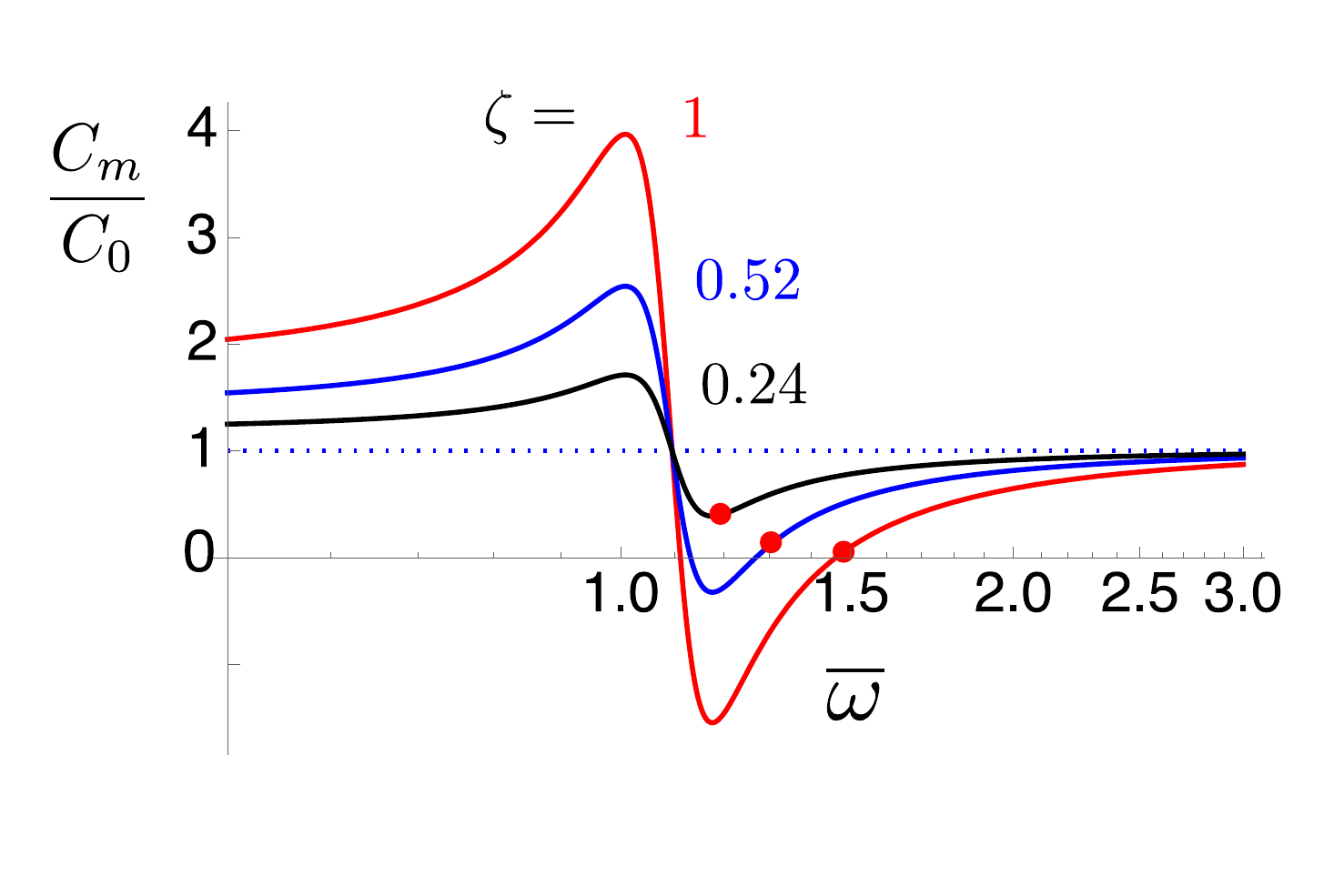}\ \ \includegraphics[width=0.45\linewidth]{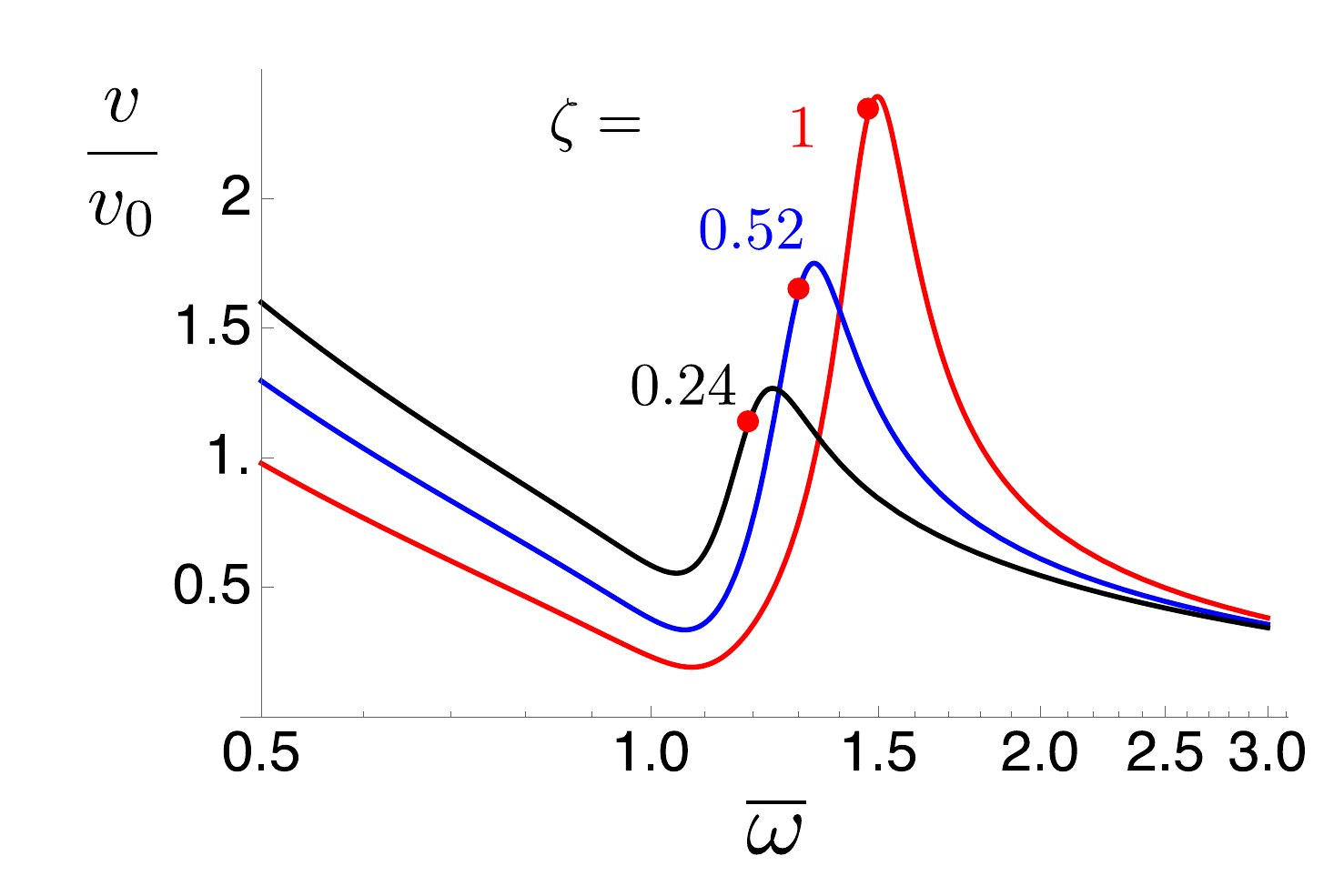}
\caption{\small{\brev Frequency dependence of the membrane capacitance and the amplitude of the receptor potential of OHC. The frequency axis is the reduced frequency $\overline\omega(=\omega/\omega_r)$, normalized by the frequency $\omega_r$ of mechanical resonance. A: Motile charge produces negative capacitance near resonance frequency, eliminating the structural membrane capacitance $C_0$. At the frequencies of maximal power generation by an OHC, which are marked by filled red circles, the total membrane capacitance is slightly positive, as indicated by Eq. \ref{eq:cap_min}. B: The receptor potential resulting from the frequency dependence of the capacitance. The traces are color coded, matching with those in A. The normalization factor $v_0$ is given by $v_0=i_0r/(\omega C_0)^2$. The traces are: $\zeta=1$ (red), $0.52$ (blue), and $0.24$ (black). \erev}}
\label{fig:negative_cap}
\end{figure}

The contribution $C_{p}$ of the movement of prestin charge to membrane capacitance can be obtained from
$C_{p}=(nq/v)Re[p]$,
where $Re[...]$ indicates the real part. This results in
\begin{align}
 C_{p}=\frac{\gamma nq^2(\alpha^2-\overline\omega^2)}{(\alpha^2-\overline\omega^2)^2+(\overline\omega/\overline\omega_\eta)^2}.
 \label{eq:Cp}
\end{align}
Notice here that $C_{p}<0$ where $\overline\omega>\alpha$ is satisfied. Since the parameter $\alpha^2$ is only a little larger than unity for the most cases, $C_{p}$ is negative for the frequency region from near the resonance frequency and higher (Fig.\ \ref{fig:negative_cap}A).

\brev
The total membrane capacitance $C_m$ is the sum of the structural capacitance $C_0$ and prestin capacitance $C_p$. What is the condition for nullifying the total capacitance $C_m$ by prestin capacitance $C_p$?

The minimum value is prestin capacitance is
\begin{align}
C_p^{min}=-\frac{\gamma nq^2 \overline\omega_\eta^2}{1+2\alpha \overline\omega_\eta},
\label{eq:Cpmin}
\end{align}
where $\gamma nq^2$ is the value of $C_p$ at $\omega\rightarrow 0$. Thus, the condition for nullifying the membrane potential $C_m$ is given by
\begin{align}
\frac{C_p(0)}{C_0}\frac{\overline\omega_\eta^2}{1+2\alpha\overline\omega_\eta}=1.
\label{eq:negative}
\end{align}
Eq.\ \ref{eq:negative} indicates that $\overline\omega_\eta(=\omega_\eta/\omega_r)$ must be relatively large. That is a natural condition for resonance. In addition, a relatively large value of prestin capacitance at the zero frequency asymptote is needed to satisfy the condition. Even with a large value of 10 for $\overline\omega_\eta$, Eq.\ \ref{eq:negative} requires $C_p(0)/C_0>0.1$.

We will come back to the issue of the membrane capacitance after a brief discussion on the mechanical power output of OHCs elicited by hair bundle stimulation.
\erev

\subsubsection*{Response to hair bundle stimulation}
The effect of hair bundle resistance $R_a$ on the membrane potential $V$ can be expressed
\begin{align}
 (e_{ec}-V)/R_a=(V-e_K)/R_m+C_0\; dV/dt-nq\; dP/dt,
 \label{eq:Ra}
\end{align}
where $e_{ec}$ is the endocochlear potential, $e_K$ is the resting potential of OHC, which is primarily determined by K$^+$ conductance, and $R_a$ hair bundle conductance. 

If we assume that stimulation at the hair bundle is periodic with angular frequency $\omega$, introducing the time independent component $R_0$ and the relative amplitude $r$ of the hair bundle resistance $R_a$, we obtain
\begin{align}
 -i_0 r=(\sigma+i\omega C_0)v-i\omega nqp.
 \label{eq:recPot}
\end{align}
Here $i_0=(e_{ec}-e_K)/(R_0+R_m)$ is  the steady state current and $\sigma=1/R_0+1/R_m$ the steady state conductance, which can be dropped for high frequency stimulation because it is smaller than $\omega C_0$, the structural capacitance.

From the transformed equation of motion Eq.\ \ref{eq:p} and equation for receptor potential Eq.\ \ref{eq:recPot}, $p$ can be expressed as a linear function of $r$.
\begin{align}
 p(-\overline\omega^2+i\overline\omega/\overline\omega_\eta+\alpha^2+\zeta)=-\frac{\gamma i_0q}{\omega C_0}\; r,
 \label{eq:p}
\end{align}
where $\zeta=\gamma nq^2/C_0$, which is the ratio of prestin capacitance at the zero-frequency limit to the structural capacitance. The equation shows that the damping term $1/\overline\omega_\eta=\omega_r/\omega_\eta$ must be small not to be overdamped. Notice also that $|p|$ is inversely proportional to $C_0$.

To evaluate the power output of the OHC, only the work against drag during a half-cycle is $E_d=(1/2)\eta\omega k_s^2|anp|^2/(k_s+K_e)^2$.  The work against elastic load does not need to be considered because it is recovered during every cycle. Power output is $Wd=(\omega/\pi)E_d$. Since $p$ is inversely proportional to the structural capacitance $C_0$, small $C_0$ is advantageous for power output.

\subsubsection*{Membrane capacitance}

Under the condition of maximal power output, the total membrane capacitance can take an approximate form \cite{Iwasa2017}
\begin{align}\nonumber
C_m&=C_0+C_{p} \\
&\approx C_0\;\frac{2\alpha^2+\zeta}{2\zeta^2}\left(\frac{\omega_r}{\omega_\eta}\right)^2
\label{eq:cap_min}
\end{align}
Notice that the membrane capacitance is proportional to $C_0$ and is small if $\omega_r/\omega_\eta>1$ (Fig.\ \ref{fig:negative_cap}A).

The diminishing value of the membrane capacitance near the resonance frequency solves the RC time constant problem in OHCs. Therefore, the receptor potential of OHCs in this frequency range is not attenuated \brev (Fig. \ref{fig:negative_cap}B) at frequencies where the membrane capacitance is small. \erev Enhancement is in the amplitude of the motion associated with OHCs. This, in turn, results in tonotopic stimulation of IHCs, which generates significant rise in the DC component of their receptor potential at high frequencies.

 \section*{Discussion}

The RC time constant problem for high frequency hearing is shared by hair cells of all ears. The solution differs in avian ears and mammalian ears. 

\subsection*{Avian hair cells}

Electrical resonance, which functions as the tuning mechanism in non-mammalian hair cells, are subjected to the frequency limit, which is imposed by the low-pass RC filter of their intrinsic electric circuit, which is determined by the membrane capacitance and membrane conductance. 


\subsubsection*{Membrane area}
SHCs occupy larger areas than THCs do on the basilar papilla. Among THCs, basal hair cells take larger areas than apical cells do \cite{Hirokawa1978,gle-man2000}. Since basal cells are shorter than apical cells, the comparison of their total membrane area is not as certain. Crude estimates from published cell outlines \cite{gle-man2000}) can be made assuming cylindrical shape for apical cells and top-truncated conical shape for basal cells.  Such an attempt leads to somewhat (about 10 to 30\%) larger membrane areas for basal cells. However, measured values for the membrane capacitance are about the same, \brev suggesting the significance of hair bundle capacitance \cite{Tan2013}. \erev

Therefore, the range of the RC roll-off frequency that enables the auditory range of chickens must be accomplished by about 20-fold decrease in the membrane resistance $R_m$ from apical to basal hair cells as described in the earlier section.
The low membrane resistance $R_m$ in basal cells may require a larger metabolic rate to operate. In addition, electrical resonance involves oscillatory intracellular Ca$^{2+}$ concentrations, which require Ca$^{2+}$ removal. This metabolic constraint is consistent with the finding that the predominant metabolism of basal hair cells is oxidative as opposed to glycolytic in apical hair cells that operate at lower frequencies \cite{Mann2021}. 

\subsubsection*{Other birds}
Of avian ears, two groups, songbirds and owls, show auditory ranges and sensitivities, which are different from chickens \cite{Dooling2000}. These differences are reflected in the hair cell morphology. In barn owls, which have much larger auditory range and sensitivity, basal hair cells are much smaller than apical hair cells \cite{Fischer1994}. Nonetheless, the membrane resistance also plays the dominant role in determining the RC roll-off frequency because the difference in the membrane area is much less than the auditory range. The ratio of the surface area of the most apical THCs to the most basal hair cells is about 5:1 if the surface area can be estimated from the cell outline assuming axial symmetry (Fig.\ 2 of Ref.\ \cite{Fischer1994}). The auditory range is from 0.1 kHz to 9 kHz \cite{Konishi1984,Fischer1994}, much wider than the area ratio can accommodate.

\subsection*{Mammalian OHC}

The traveling wave proceeds from the basal end toward the apical end of the cochlea. This means energy flow accompanied by the traveling wave may contribute to the local energy balance in general, but energy influx is absent in the basal part where the wave initiates. In other words, the condition for local energy balance involving resonance can be considered as a necessary condition for an upper bound of the auditory frequency as described above.

\subsubsection*{Piezoelectric resonance}
Mammalian OHCs can overcome the limitation imposed by the low-pass RC filter by significantly reducing the membrane capacitance with piezoelectric resonance, resulting from the combination of the piezoelectric property of prestin in OHCs and mechanical resonance in the cochlea \cite{Iwasa2017,Iwasa2021}. The residual membrane capacitance is proportional to the structural membrane capacitance $C_0$, and the maximal power output of OHCs is inversely proportional to $C_0$. Therefore the surface area must be small to operate at high frequencies consistent with morphological observations \cite{Pujol1992}.

Analogous to avian hair cells, basal OHCs do have smaller basolateral resistance \cite{Johnson2011}. This finding appears as if lower membrane resistance $R_m$ can elevate the RC roll-off frequency. It should be noted, however, that a reduction of the membrane resistance $R_m$ only reduces the receptor potential in OHCs unlike avian hair cells. This appears to match the larger hair bundle conductance of the basal cells \cite{Beurg2015}. 

\brev
\subsection*{The role of prestin}
It is of interest that prestin exists both in mammalian OHCs and avian SHCs and generates force that can sharpen the tuning of these ears in response to the receptor potential of these cell types, even though the localization of this membrane protein is different. In OHCs this protein is in the lateral membrane, and in SHCs it is associated with hair bundles  \cite{Beurg2013}.

Another major difference is in its significance to the membrane capacitance. Whereas the contribution of prestin capacitance can be as large as the structural membrane capacitance in OHCs, its contribution to the membrane capacitance is up to 2\% of the structural capacitance \cite{Beurg2013}. Even though avian prestin can be associated with mechanical resonance, the negative capacitance created by prestin is too small to affect the membrane capacitance (See Eq.\ \ref{eq:negative}).
For this reason, avian prestin is not expected to have any significant effect in affecting the RC time constant of the SHCs, quite unlike in OHCs.
Nonetheless, the role of prestin in avian SHCs in providing feedback could be indicative of the evolutionary pathway to mammalian prestin in OHCs \cite{Beurg2013}.
\erev

\subsection*{The auditory ranges}
Elevating the RC roll-off frequency to the high end of the auditory frequencies by reducing the membrane resistance appears to have a limit. Even for chickens, the auditory range of which is up to 4 kHz, the reception of the upper end \brev needs to be aided by extracellular mechanism. The tonotopic tilting of the tectorial membrane could be associated to such a mechanism. \erev For mammalian ears, the auditory range of which can extend to 80 kHz and beyond, elevating the RC roll-off frequency by decreasing membrane resistance does not appear practical. The only practical way of solving the RC time constant problem is to reduce the membrane capacitance. That can be accomplished by piezoelectric resonance. 

\section*{Acknowledgments}
I thank Drs.\ Catherine Weisz and Richard Chadwick of NIDCD for comments. This research was supported in part by the Intramural Research Program of the NIH, NIDCD.


\begin{thebibliography}{59}
\providecommand{\url}[1]{\texttt{#1}}
\providecommand{\urlprefix}{ }

\bibitem[Ehret(1976)]{Ehret1976}
Ehret, G., 1976.
\newblock Development of absolute auditor thresholds in the house mouse ({M}us
  musculus).
\newblock \emph{J. Am. Audiol. Soc.} 1:179--184.

\bibitem[Rebillard and Rubel(1981)]{Rebillard1981}
Rebillard, G., and E.~W. Rubel, 1981.
\newblock Electrophysiological study of the maturation of auditory responses
  from the inner ear of the chick.
\newblock \emph{Brain Res.} 229:15--23.

\bibitem[Sullivan and Konishi(1987)]{Konishi1984}
Sullivan, W.~E., and M.~Konishi, 1987.
\newblock Segregation of stimulus phase and intensity coding in the cochlear
  nucleus of the barn owl.
\newblock \emph{J. Neurosci.} 4:1787--1799.

\bibitem[Brownell et~al.(1985)Brownell, Bader, Bertrand, and
  Ribaupierre]{bbbr1985}
Brownell, W., C.~Bader, D.~Bertrand, and Y.~Ribaupierre, 1985.
\newblock Evoked mechanical responses of isolated outer hair cells.
\newblock \emph{Science} 227:194--196.

\bibitem[Ashmore(1987)]{a1987}
Ashmore, J.~F., 1987.
\newblock A fast motile response in guinea-pig outer hair cells: the molecular
  basis of the cochlear amplifier.
\newblock \emph{J. Physiol.} 388:323--347.

\bibitem[Dallos et~al.(2008)Dallos, Wu, Cheatham, Gao, Zheng, Anderson, Jia,
  Wang, Cheng, Sengupta, He, and Zuo]{Dallos2008}
Dallos, P., X.~Wu, M.~A. Cheatham, J.~Gao, J.~Zheng, C.~T. Anderson, S.~Jia,
  X.~Wang, W.~H.~Y. Cheng, S.~Sengupta, D.~Z.~Z. He, and J.~Zuo, 2008.
\newblock Prestin-based outer hair cell motility is necessary for mammalian
  cochlear amplification.
\newblock \emph{Neuron} 58:333--339.

\bibitem[Housley and Ashmore(1992)]{ha1992}
Housley, G.~D., and J.~F. Ashmore, 1992.
\newblock Ionic currents of outer hair cells isolated from the guinea-pig
  cochlea.
\newblock \emph{J. Physiol.} 448:73--98.

\bibitem[Dallos and Evans(1995)]{Dallos1995}
Dallos, P., and B.~N. Evans, 1995.
\newblock High-frequency outer hair cell motility: corrections and addendum.
\newblock \emph{Science} 268:1420--1421.

\bibitem[Mistr\'ik et~al.(2009)Mistr\'ik, Mullaley, Mammano, and
  Ashmore]{Mistrik2009}
Mistr\'ik, P., C.~Mullaley, F.~Mammano, and J.~Ashmore, 2009.
\newblock Three-dimensional current flow in a large-scale model of the cochlea
  and the mechanism of amplification of sound.
\newblock \emph{J R Soc Interface} 6:279--291.

\bibitem[Spector et~al.(2003)Spector, Brownell, and Popel]{Spector2003}
Spector, A.~A., W.~E. Brownell, and A.~S. Popel, 2003.
\newblock Effect of outer hair cell piezoelectricity on high-frequency receptor
  potentials.
\newblock \emph{J Acoust Soc Am} 113:453--461.

\bibitem[Rabbitt et~al.(2009)Rabbitt, Clifford, Breneman, Farrell, and
  Brownell]{Rabbitt2009}
Rabbitt, R.~D., S.~Clifford, K.~D. Breneman, B.~Farrell, and W.~E. Brownell,
  2009.
\newblock Power efficiency of outer hair cell somatic electromotility.
\newblock \emph{PLoS Comput. Biol.} 5:e1000444.

\bibitem[Johnson et~al.(2011)Johnson, Beurg, Marcotti, and
  Fettiplace]{Johnson2011}
Johnson, S.~L., M.~Beurg, W.~Marcotti, and R.~Fettiplace, 2011.
\newblock Prestin-driven cochlear amplification is not limited by the outer
  hair cell membrane time constant.
\newblock \emph{Neuron} 70:1143--1154.

\bibitem[K\"oppl(2015)]{Koeppl2015}
K\"oppl, C., 2015.
\newblock Chapter 6. {A}vian hearing.
\newblock \emph{In} C.~G. Sacnes, editor, Sturkie's Avian Physiology (6th
  edition), Academic Press, 71--87.

\bibitem[Fettiplace(2020)]{Fettiplace2020}
Fettiplace, R., 2020.
\newblock Diverse Mechanisms of Sound Frequency Discrimination in the
  Vertebrate Cochlea.
\newblock \emph{Trends Neurosci.} 43:88--102.

\bibitem[Santos-Sacchi(1991)]{s1991}
Santos-Sacchi, J., 1991.
\newblock Reversible inhibition of voltage-dependent outer hair cell motility
  and capacitance.
\newblock \emph{J. Neurophysiol.} 11:3096--3110.

\bibitem[Adachi and Iwasa(1999)]{ai1999}
Adachi, M., and K.~H. Iwasa, 1999.
\newblock Electrically driven motor in the outer hair cell: Effect of a
  mechanical constraint.
\newblock \emph{Proc. Natl. Acad. Sci. USA} 96:7244--7249.

\bibitem[Iwasa(2017)]{Iwasa2017}
Iwasa, K.~H., 2017.
\newblock Negative membrane capacitance of outer hair cells: electromechanical
  coupling near resonance.
\newblock \emph{Sci. Rep.} 7:12118.

\bibitem[Iwasa(2021)]{Iwasa2021}
Iwasa, K.~H., 2021.
\newblock Kinetic membrane model of outer hair cells.
\newblock \emph{Biophys. J.} 120:122--132.

\bibitem[Crawford and Fettiplace(1981)]{CrawFett1981}
Crawford, A.~C., and R.~Fettiplace, 1981.
\newblock An electrical tuning mechanism in turtle cochlear hair cells.
\newblock \emph{J. Physiol.} 312:377--412.

\bibitem[Hudspeth and Lewis(1988)]{HudLew1988}
Hudspeth, A.~J., and R.~S. Lewis, 1988.
\newblock A model for electrical resonance and frequency tuning in saccular
  hair cells of the bull-frog, \emph{{R}ana {C}astebeliana}.
\newblock \emph{J Physiol.} 400:275--297.

\bibitem[Fuchs et~al.(1988)Fuchs, Nagai, and Evans]{Fuchs1988}
Fuchs, P.~A., T.~Nagai, and M.~G. Evans, 1988.
\newblock Electrical tuning in hair cells isolated from the chick cochlea.
\newblock \emph{J Neurosci} 8:2460--2467.

\bibitem[Wu et~al.(1995)Wu, Art, Goodman, and Fettiplace]{WuFett1995}
Wu, Y.-C., J.~J. Art, M.~B. Goodman, and R.~Fettiplace, 1995.
\newblock A kinetic description of the Calcium-activated potassium channel and
  its application to electrical tuning of hair cells.
\newblock \emph{Prog. Biophys. Molec Biol.} 63:131--158.

\bibitem[Jones et~al.(1998)Jones, Laus, and Fettiplace]{JonesFett1998}
Jones, E. M.~C., C.~Laus, and R.~Fettiplace, 1998.
\newblock Identification of {C}a$^{2+}$-activated {K}$^+$ channel splice
  variants and their distribution in the turtle cochlea.
\newblock \emph{Proc. Roy. Soc. Lond. {B}} 265:685--692.

\bibitem[Mauro et~al.(1970)Mauro, Conti, Dodge, and Schor]{Mauro1970}
Mauro, A., F.~Conti, F.~Dodge, and R.~Schor, 1970.
\newblock Subthreshold behavior and phenomenological impedance of the squid
  giant axon.
\newblock \emph{J. Gen. Physiol.} 55:497--523.

\bibitem[Gleich and Manley(2000{\natexlab{a}})]{Gleich2000}
Gleich, O., and G.~A. Manley, 2000.
\newblock The hearing organ of birds and crocodilia.
\newblock \emph{In} R.~J. Dooling, R.~R. Faye, and A.~N. Popper, editors,
  Comparative Hearing: Birds and Reptiles, Springer, New York, NY, 70--138.

\bibitem[Tan et~al.(2013{\natexlab{a}})Tan, Beurg, Hackney, Mahendrasingam, and
  Fettiplace]{TanFett2013}
Tan, X., M.~Beurg, C.~Hackney, S.~Mahendrasingam, and R.~Fettiplace, 2013.
\newblock Electrical tuning and transduction in short hair cells of the chicken
  auditory papilla.
\newblock \emph{J. Neurophysiol.} 109:2007 -- 2020.

\bibitem[Gummer et~al.(1987)Gummer, Smolders, and Klinke]{Gummer1987}
Gummer, A.~W., J.~W. Smolders, and R.~Klinke, 1987.
\newblock Basilar membrane motion in the pigeon measured with the
  {M}\"{o}ssbauer technique.
\newblock \emph{Hear Res} 29:63--92.

\bibitem[Beurg et~al.(2013)Beurg, Tan, and Fettiplace]{Beurg2013}
Beurg, M., X.~Tan, and R.~Fettiplace, 2013.
\newblock A prestin motor in chicken auditory hair cells: active force
  generation in a nonmammalian species.
\newblock \emph{Neuron} 79:69--81.

\bibitem[Steele(1997)]{Steele1997}
Steele, C.~R., 1997.
\newblock Three-dimensional mechanical modeling of the cochlea.
\newblock \emph{In} E.~R. Lewis, G.~R. Long, R.~F. Lyon, P.~M. Narins, C.~R.
  Steele, and E.~Hecht-Poinar, editors, Diversity in Auditory Mechanics, World
  Scientific, Singapore, 455--461.

\bibitem[Choe et~al.(1998)Choe, Magnasco, and Hudspeth]{cho-hud1998}
Choe, Y., M.~O. Magnasco, and A.~J. Hudspeth, 1998.
\newblock A model for amplification of hair-bundle motion by cyclical binding
  of {C}a$^{2+}$ to mechanoelectrical-transduction channel.
\newblock \emph{Proc. Natl. Acad. Sci. USA} 95:15321--15326.

\bibitem[Tinevez et~al.(2007)Tinevez, J\"{u}licher, and Martin]{Tinevez2007}
Tinevez, J.-Y., F.~J\"{u}licher, and P.~Martin, 2007.
\newblock Unifying the various incarnations of active hair-bundle motility by
  the vertebrate hair cell.
\newblock \emph{Biophys J} 93:4053--4067.

\bibitem[Sul and Iwasa(2009)]{Sul2009c}
Sul, B., and K.~H. Iwasa, 2009.
\newblock Amplifying effect of a release mechanism for fast adaptation in the
  hair bundle.
\newblock \emph{J Acoust Soc Am} 126:4--6.

\bibitem[K\"oppl et~al.(1998)K\"oppl, Gleich, Schwabedissen, Siegl, and
  Manley]{Koeppl1998}
K\"oppl, C., O.~Gleich, G.~Schwabedissen, E.~Siegl, and G.~A. Manley, 1998.
\newblock Fine structure of the basilar papilla of the emu: implications for
  the evolution of avian hair-cell types.
\newblock \emph{Hear Res} 126:99--112.

\bibitem[Gleich and Manley(2000{\natexlab{b}})]{gle-man2000}
Gleich, O., and G.~A. Manley, 2000.
\newblock The hearing organ of birds and crocodilia.
\newblock \emph{In} R.~J. Dooling, R.~R. Fay, and P.~A. N., editors,
  Comparative Hearing: Birds and Reptiles, Springer, New York, 70--138.

\bibitem[Iwasa and Ricci(2015)]{iwasa2014}
Iwasa, K.~H., and A.~J. Ricci, 2015.
\newblock The avian tectorial membrane: Why is it tapered.
\newblock \emph{In} K.~D. Karavitaki, and D.~P. Corey, editors, Mechanics of
  Hearing: Protein to Perception, American Institute of Physics, Melville, NY,
  08005.1--08005.6.

\bibitem[Smith et~al.(1985)Smith, Konishi, and Schuff]{Smith1985}
Smith, C.~A., M.~Konishi, and N.~Schuff, 1985.
\newblock Structure of the barn owl's (Tyto alba) inner ear.
\newblock \emph{Hear Res} 17:237--247.

\bibitem[Xia et~al.(2016)Xia, Liu, Raphael, Applegate, and Oghalai]{XiaOgh2016}
Xia, A., X.~Liu, P.~D. Raphael, B.~E. Applegate, and J.~S. Oghalai, 2016.
\newblock Hair cell force generation does not amplify or tune vibrations within
  the chicken basilar papilla.
\newblock \emph{Nature Commun.} 12133.

\bibitem[Goodyear et~al.(1994)Goodyear, Holley, and Richardson]{Goodyear1994}
Goodyear, R., M.~Holley, and G.~Richardson, 1994.
\newblock Visualisation of domains in the avian tectorial and otolithic
  membranes with monoclonal antibodies.
\newblock \emph{Hear Res} 80:93--104.

\bibitem[B\'{e}k\'{e}sy(1952)]{bekesy1952}
B\'{e}k\'{e}sy, G.~v., 1952.
\newblock $\textrm{DC}$ resting potentials inside the cochlear partition.
\newblock \emph{J. Acoust. Soc. Am.} 24:72--76.

\bibitem[O~Maoil\'eidigh and Hudspeth(2013)]{OMaoileidigh2013}
O~Maoil\'eidigh, D., and A.~J. Hudspeth, 2013.
\newblock Effects of cochlear loading on the motility of active outer hair
  cells.
\newblock \emph{Proc. Natl. Acad. Sci. USA} 110:5474--5479.

\bibitem[Iwasa(2016)]{Iwasa2016}
Iwasa, K.~H., 2016.
\newblock Energy Output from a Single Outer Hair Cell.
\newblock \emph{Biophys. J.} 111:2500--2511.

\bibitem[Liu et~al.(2017)Liu, Gracewski, and Nam]{Liu2017}
Liu, Y., S.~M. Gracewski, and J.-H. Nam, 2017.
\newblock Two passive mechanical conditions modulate power generation by the
  outer hair cells.
\newblock \emph{PLoS Comp Biol} 13:e1005701.

\bibitem[Rabbitt(2020)]{Rabbitt2020}
Rabbitt, R.~D., 2020.
\newblock The cochlear outer hair cell speed paradox.
\newblock \emph{Proc Natl Acad Sci U S A} 117:21880--21888.

\bibitem[Cole(1968)]{cole}
Cole, K.~S., 1968.
\newblock Membranes, ions, and impulses.
\newblock University of California Press, Berkely, CA.

\bibitem[Santos-Sacchi et~al.(1998)Santos-Sacchi, Kakehata, and
  Takahashi]{Santos-Sacchi1998a}
Santos-Sacchi, J., S.~Kakehata, and S.~Takahashi, 1998.
\newblock Effects of membrane potential on the voltage dependence of
  motility-related charge in outer hair cells of the guinea-pig.
\newblock \emph{J Physiol} 510 ( Pt 1):225--235.

\bibitem[Zheng et~al.(2000)Zheng, Shen, He, Long, Madison, and
  Dallos]{zshlmd2000}
Zheng, J., W.~Shen, D.~Z.-Z. He, K.~B. Long, L.~D. Madison, and P.~Dallos,
  2000.
\newblock Prestin is the motor protein of cochlear outer hair cells.
\newblock \emph{Nature} 405:149--155.

\bibitem[Gale and Ashmore(1997)]{ga1997}
Gale, J.~E., and J.~F. Ashmore, 1997.
\newblock An intrinsic frequency limit to the cochlear amplifier.
\newblock \emph{Nature} 389:63--66.

\bibitem[Iwasa(1993)]{i1993}
Iwasa, K.~H., 1993.
\newblock Effect of stress on the membrane capacitance of the auditory outer
  hair cell.
\newblock \emph{Biophys. J.} 65:492--498.

\bibitem[Gale and Ashmore(1994)]{ga1994}
Gale, J.~E., and J.~F. Ashmore, 1994.
\newblock Charge displacement induced by rapid stretch in the basolateral
  membrane of the guinea-pig outer hair cell.
\newblock \emph{Proc. Roy. Soc. (Lond.) B Biol. Sci.} 255:233--249.

\bibitem[Dong et~al.(2002)Dong, Ospeck, and Iwasa]{doi2002}
Dong, X.~X., M.~Ospeck, and K.~H. Iwasa, 2002.
\newblock Piezoelectric reciprocal relationship of the membrane motor in the
  cochlear outer hair cell.
\newblock \emph{Biophys. J.} 82:1254--1259.

\bibitem[Mountain and Hubbard(1994)]{mh1994}
Mountain, D.~C., and A.~E. Hubbard, 1994.
\newblock A piezoelectric model of outer hair cell function.
\newblock \emph{J. Acoust. Soc. Am.} 95:350--354.

\bibitem[Weitzel et~al.(2003)Weitzel, Tasker, and Brownell]{Weitzel2003}
Weitzel, E.~K., R.~Tasker, and W.~E. Brownell, 2003.
\newblock Outer hair cell piezoelectricity: Frequency response enhancement and
  resonance behavior.
\newblock \emph{J. Acoust. Soc. Am.} 114:1462--1466.

\bibitem[Hirokawa(1978)]{Hirokawa1978}
Hirokawa, N., 1978.
\newblock The ultrastructure of the basilar papilla of the chick.
\newblock \emph{J Comp Neurol} 181:361--374.

\bibitem[Tan et~al.(2013{\natexlab{b}})Tan, Beurg, Hackney, Mahendrasingam, and
  Fettiplace]{Tan2013}
Tan, X., M.~Beurg, C.~Hackney, S.~Mahendrasingam, and R.~Fettiplace, 2013.
\newblock Electrical tuning and transduction in short hair cells of the chicken
  auditory papilla.
\newblock \emph{J Neurophysiol} 109:2007--2020.

\bibitem[Mann()]{Mann2021}
Mann, Z.
\newblock personal communication.

\bibitem[Dooling et~al.(2000)Dooling, Lohr, and Dent]{Dooling2000}
Dooling, R.~J., B.~Lohr, and M.~L. Dent, 2000.
\newblock Hearing in birds and reptiles.
\newblock \emph{In} R.~J. Dooling, R.~R. Faye, and A.~N. Popper, editors,
  Comparative Hearing: Birds and Reptiles, Springer, New York, NY, 308--359.

\bibitem[Fischer(1994)]{Fischer1994}
Fischer, F.~P., 1994.
\newblock Quantitative {TEM} analysis of the barn owl basilar papilla.
\newblock \emph{Hearing Res.} 73:1--15.

\bibitem[Pujol et~al.(1992)Pujol, Lenoir, Ladrech, Tribillac, and
  Rebillard]{Pujol1992}
Pujol, R., M.~Lenoir, S.~Ladrech, F.~Tribillac, and G.~Rebillard, 1992.
\newblock Correlation between the length of outer hair cells and the frequency
  coding of the cochlea.
\newblock \emph{In} Y.~Cazals, L.~Demany, and K.~Horner, editors, Auditory
  Physiology and Perception, Pergamon Press, 45--52.

\bibitem[Beurg et~al.(2015)Beurg, Xiong, Zhao, M\"uller, and
  Fettiplace]{Beurg2015}
Beurg, M., W.~Xiong, B.~Zhao, U.~M\"uller, and R.~Fettiplace, 2015.
\newblock Subunit determination of the conductance of hair-cell
  mechanotransducer channels.
\newblock \emph{Proc. Natl. Acad. Sci. USA} 112:1589--1594.

\end{thebibliography}

\end{document}